\documentclass{aa}
\renewcommand\makeLineNumber{}

\usepackage{graphicx}
\usepackage[varg]{txfonts}
\usepackage[linkcolor=magenta,citecolor=blue,colorlinks=true]{hyperref}
\begin{document}

   \title{Detection probability of light compact binary mergers in future observing runs of the current ground-based gravitational wave detector network}
   \titlerunning{BNS and BHNS merger detection probability forecast}

   \author{{Om Sharan} Salafia
          \inst{1,2}
          }

   \institute{INAF -- Osservatorio Astronomico di Brera, via Brera 28, I-20121 Milano (MI), Italy
         \and
             INFN, Sezione di Milano-Bicocca, Piazza della Scienza 2, I-20126 Milano (MI), Italy\\
             }

   \date{\today}

% \abstract{}{}{}{}{} 
% 5 {} token are mandatory
 
  \abstract 
   {With no binary neutron star (BNS) merger detected yet during the fourth observing run (O4) of the LIGO-Virgo-KAGRA (LVK) gravitational wave (GW) detector network, despite the time-volume (VT) surveyed with respect to the end of O3 increased by more than a factor of three, a pressing question is how likely the detection of at least one BNS merger is in the remainder of the run. I present here a simple and general method to address such a question, which constitutes the basis for the predictions that have been presented in the LVK Public Alerts User Guide during the hiatus between the O4a and O4b parts of the run. The method, which can be applied to neutron star - black hole (NSBH) mergers as well, is based on simple Poisson statistics and on an estimate of the ratio of the VT span by the future run to that span by previous runs. An attractive advantage of this method is that its predictions are independent from the mass distribution of the merging compact binaries, which is very uncertain at the present moment. The results, not surprisingly, show that the most likely outcome of the final part of O4 is the absence of any BNS merger detection. Still, the probability of a non-zero number of detections is 34-46\%. For NSBH mergers, the probability of at least one additional detection is 64-71\%. The prospects for the next observing run O5 are more promising, with predicted numbers $N_\mathrm{BNS,O5}=28_{-21}^{+44}$, and the NSBH detections to be $N_\mathrm{NSBH,O5}=65_{-38}^{+61}$ (median and 90\% symmetric credible range), based on the current LVK detector target sensitivities for the run. The calculations presented here also lead to an update of the LVK local BNS merger rate density estimate that accounts for the absence of BNS merger detections in O4 so far, that reads $2.8\,\mathrm{Gpc^{-3}\,yr^{-1}}\leq R_0\leq 480\,\mathrm{Gpc^{-3}\,yr^{-1}}$.}

   \keywords{gravitational waves, stars: neutron, stars: black holes, methods: statistical}

   \maketitle

%
%-------------------------------------------------------------------

\section{Introduction}

Binary neutron star (BNS) mergers are one of the main sources of gravitational waves (GW) in the frequency range of sensitivity of the current ground-based GW detectors, such as the two detectors in the Advanced Laser Interferometer Gravitational wave Observatory (LIGO, \citealt{LIGO2015}), the Advanced Virgo \citep{Virgo2015} detector, and the KAGRA \citep{KAGRA2012} detector. These sources are of particular interest because of their multi-messenger nature: in addition to GW emitted during the inspiral, merger and post-merger phases, BNS coalescences also produce kilonovae \citep{Li1998,Metzger2020LRR} and non-thermal emission related to the launch of a relativistic jet \citep[e.g.][]{Eichler1989,Lazzati2017,Nakar2020}. As demonstrated by the spectacular GW170817 event \citep{Abbott2017_GW170817,Abbott2017_MM,Margutti2021},  observations of these multiple messengers can have a tremendous impact on several branches of physics, including fundamental physics \citep[e.g.][]{Abbott2019_GWtest,Baker2017,Creminelli2017}, nuclear physics \citep[e.g.][]{Abbott2018_EoS}, cosmology \citep[e.g.][]{Abbott2017_H0}, the synthesis of heavy elements \citep[e.g.][]{Coulter2017,Pian2017,Kasen2017,Kajino2019}, the physics of gamma-ray bursts and their jets \citep[e.g.][]{Abbott2017_GW_and_GRB,Savchenko2017,Kasliwal2017,Mooley2018,Ghirlanda2019}, massive binary stellar evolution \citep[e.g.][]{Kruckow2018,Mapelli2018}, to name only a few.

The transformative potential of multi-messenger observations of these sources must confront with the very challenging nature of their electromagnetic follow-up, due to the faintness of the electromagnetic emission combined with the poor GW localization (see e.g. \citealt{Nicholl2025}). In the last several years, the international transient astronomy community put a large effort in order to be ready to take this challenge. Such organizational effort, which includes allocating human resources and applying for observing time at the best astronomical facilities worldwide, is guided by, and gauged on, predicitions of the expected detection rate of such events. Most such predictions for the current observing run O4 of the LIGO-Virgo-KAGRA (LVK) network proved to be rather optimistic \citep[see][who provide a convenient summary of many predictions from the literature in their discussion section]{Colombo2022}, and even the official LVK predictions (available on the LVK Public Alerts User Guide web page\footnote{\url{https://emfollow.docs.ligo.org/userguide/capabilities.html}} and based on \citealt{Petrov2022}) promised tens of BNS merger public alerts to be delivered by the GW detector network during O4.

In this work, I present a method to calculate the probability of future BNS (and neutron star - black hole, NSBH) merger detections by the LVK network that is independent of the uncertain mass distribution of the merging binaries, and that uses only the information on the past number of detections, combined with an estimate of the ratio between the sensitivity of the target run with respect to the previous runs that requires only publicly-available information as input. Using this method, I provide updated predictions for the probability of one or more BNS and NSBH detections in the remainder of O4, and in the next observing run O5.

\section{Poisson probability informed by previous occurrences of a rare event}

I derive here the expression that I used in order to forecast the probability of detecting GW from new BNS and BHNS events in future observing runs. After carrying out the derivation, I was informed that it is essentially identical to that presented in Appendix B of \citet{Ray2023}, and that the result coincides with Eq.\ 42 in \citet{Essick2023} in a particular case.

Let $N$ be the a number of occurrences of a rare event over a period of time $T$, and let $\lambda$ be the expected number of events, that is, the average occurrence rate multiplied by $T$. The probability of $N$ given $\lambda$ is the Poisson probability
\begin{equation}
 p(N\,|\,\lambda) = \frac{\lambda^N\exp(-\lambda)}{N!}.
\label{eq:Poisson}
 \end{equation}
Now let $N'$ be a number of previously observed events over a different time period $T'$, over which the expected number of events was $\lambda'$, with $\mathcal{C}=\lambda/\lambda'$. The posterior probability of $\lambda'$ given $N'$ can be written through Bayes' theorem as
\begin{equation}
 p(\lambda'\,|\,N') = \frac{p(N'\,|\,\lambda')\pi(\lambda')}{p(N')}=\frac{{\lambda^\prime}^{N^\prime}\exp(-\lambda')}{N'!}\frac{\pi(\lambda')}{p(N')},
\end{equation}
where $\pi(\lambda')$ is the prior probability of $\lambda'$. We opt to parametrize this as
\begin{equation}
\pi(\lambda')=\left(\frac{p(N')N'!}{\Gamma(N'+1-\alpha)}\right)\lambda^{-\alpha},
\end{equation}
where $\Gamma(x)$ is the Gamma function and the factor in parentheses ensures the correct normalization of the posterior. With this definition, $\alpha=0$ corresponds to a uniform prior, $\alpha=1/2$ to the Jeffreys prior for the Poisson probability scale parameter, and $\alpha=1$ to a prior that is uniform in the logarithm. Since these are the most common choices for a un-informative prior on this parameter, I limit here the discussion to these cases.

The above definitions allow us to derive the posterior probability on $N$ given the previously observed number of events $N'$, as follows. The starting point is
\begin{equation}
\begin{split}
& p(N\,|\,N') = \int p(N\,|\,\lambda) p(\lambda\,|\,N')\,\mathrm{d}\lambda =\\
& = \int p(N\,|\,\lambda) \int p(\lambda\,|\,\lambda') p(\lambda'\,|\,N')\,\mathrm{d}\lambda'\,\mathrm{d}\lambda.
\end{split}
\end{equation}
Noting that $p(\lambda\,|\,\lambda')=\delta(\lambda - \mathcal{C}\lambda')$, where $\delta$ is the Dirac delta, this leads to
\begin{equation}
 p(N\,|\,N') = \frac{\mathcal{C}^{\alpha-N'-1}}{\Gamma(N'+1-\alpha)}\int p(N\,|\,\lambda)\exp\left(-\frac{\lambda}{\mathcal{C}}\right)\lambda^{N'-\alpha}\,\mathrm{d}\lambda.
\end{equation}
Substituting Eq.\,\ref{eq:Poisson} in the above expression, carrying out the integral, and using $N! = \Gamma(N+1)$, we arrive to
\begin{equation}
 p(N\,|\,N',\alpha,\mathcal{C}) = \frac{\Gamma(N+N'+1-\alpha)}{\Gamma(N+1)\Gamma(N'+1-\alpha)}\frac{\mathcal{C}^N}{(1+\mathcal{C})^{N+N'+1-\alpha}},
 \label{eq:P(N|N')}
\end{equation}
where the dependency on the prior index $\alpha$ and the expected number ratio $\mathcal{C}$ has been made explicit.

\section{Application to compact binary mergers}

If the sensitivity range of the gravitational wave detector network to the sources under consideration does not extend to large redshifts, the cosmic evolution of the population and cosmological effects can be neglected, so that the expected number of detections over an observing run of duration $T$ can be expressed simply as $\lambda=R_0 V T$, where $R_0$ is the local rate density of compact binary mergers and $V$ is the effective volume over which the network is sensitive to such sources. In this context, the ratio $\lambda/\lambda'$ is then simply the ratio of the effective sensitive time-volume of the run to that of past runs, namely
\begin{equation}
 \mathcal{C}=\frac{V T}{\sum_{l=0}^{n_\mathrm{past}-1}V_{l}T_{l}},
 \label{eq:C_simple}
\end{equation}
where $l$ runs over past observing runs, whose total number is $n_\mathrm{past}$. In the following, I describe a strategy to estimate such ratio using basic information such as the binary neutron star (BNS) ranges and duty cycles of the detectors.

\subsection{Evaluation of the effective sensitive volume for each run}

The `optimal' matched filter signal-to-noise ratio (S/N) of a single merger $\rho_\mathrm{opt}$, assuming it to be dominated by the inspiral part of the signal, depends on the chirp mass $m_\mathrm{c}=(m_1 m_2)^{3/5}/(m_1+m_2)^{1/5}$ (where $m_1$ and $m_2$ are the gravitational masses of the primary and secondary components of the binary), the luminosity distance $r$, and on the detector noise power spectral density (PSD) curve $\mathcal{S}(f)$ as a function of frequency $f$ through the integral\footnote{I neglect here a small additional dependence on the component masses and possibly on the neutron star matter equation of state, which together determine the effective inspiral cut-off frequency.} $f_{7/3}=\int \left[f^{7/3}\mathcal{S}(f)\right]^{-1}\mathrm{d}f$ \citep{Finn1993}, so that
\begin{equation}
 \rho_\mathrm{opt}\propto \frac{m_\mathrm{c}^{5/6}}{r}f_{7/3}.
 \label{eq:snropt}
\end{equation}
The actual S/N in a given detector, which I indicate with the symbol $\rho$, also depends on the source position in the detector's sky (defined e.g.\ by a pair of spherical angular coordinates $\theta$, $\phi$), its inclination $\iota$ with respect to the line of sight and its polarization angle $\psi$, all of which can be summarized into a single parameter $0\leq w\leq 1$ such that $\rho = w\rho_\mathrm{opt}$ \citep{Finn1993,Dominik2015,Chen2021}, with
\begin{equation}
 w^2 = \frac{1}{4}F_{+}^2(\theta,\phi,\psi)(1+\cos^2\iota)^2+F_\times^2(\theta,\phi,\psi)\cos^2\iota,
\end{equation}
where $F_+$ and $F_\times$ are the `antenna pattern' functions that define the dependence of the detector's sensitivity on sky position and polarization angle, which can be expressed as \citep[e.g.][]{Dhurandhar1988}
\begin{equation}
\begin{split}
 & F_+(\theta,\phi,\psi) = \frac{1}{2}(1+\cos^2\theta)\cos 2\phi \cos 2\psi - \cos\theta\sin 2\phi \sin 2\psi,\\
& F_\times(\theta,\phi,\psi) = \frac{1}{2}(1+\cos^2\theta)\cos 2\phi \cos 2\psi + \cos\theta\sin 2\phi \cos 2\psi.
\end{split}
\end{equation}
The probability distribution of $w$ for each detector is completely specified under the assumption of isotropic sky positions and binary orbital plane orientations. Since $w\leq 1$, and given the dependencies in Eq.\ \ref{eq:snropt}, for each detector there exists a `horizon' distance $d_\mathrm{h}(m_\mathrm{c})\propto m_\mathrm{c}^{5/6} f_{7/3}$ such that $\rho_\mathrm{opt}(r=d_\mathrm{h})=\rho_\mathrm{lim}$, where $\rho_\mathrm{lim}$ is a minimum SNR required for a detection. This represents the distance beyond which a binary with a chirp mass $m_\mathrm{c}$ cannot be detected. Hence, for a given binary, one can write the S/N in the $i$-th detector of a network as
\begin{equation}
 \rho_i = w_i \rho_\mathrm{lim}\frac{d_{\mathrm{h},i}}{r},
\end{equation}
and the squared `network S/N' in an $n$-detector network as
\begin{equation}
 \rho_\mathrm{net}^2 = \sum_{i=0}^{n-1} \rho_i^2 = w_0^2\rho_\mathrm{lim}^2\frac{d_{\mathrm{h},0}^2}{r^2}\left[1+\sum_{i=1}^{n-1}\left(\frac{w_i}{w_0}\right)^2\left(\frac{d_{\mathrm{h},i}}{d_{\mathrm{h},0}}\right)^2\right].
\label{eq:snrnet}
 \end{equation}

Let us now represent the detection as a threshold on the network S/N $\rho_\mathrm{net}\geq \rho_\mathrm{lim}$. In other words, let us define the detection probability of a binary merger as
\begin{equation}
 p_\mathrm{det} = \Theta\left(\rho_\mathrm{net}-\rho_\mathrm{lim}\right),
\end{equation}
where $\Theta$ is the Heaviside step function. The effective sensitive volume of the network, neglecting cosmological effects, is then obtained by integrating the detection probability over volume and over the binary orientations,
\begin{equation}
\begin{split}
 & V_\mathrm{eff}=\iiiint r^2 p_\mathrm{det}\mathrm{d}r \sin\theta\,\mathrm{d}\theta\,\mathrm{d}\phi \,\frac{\sin\iota\,\mathrm{d}\iota }{2}\frac{\mathrm{d}\psi}{2\pi}=\\
 & 4\pi d_{\mathrm{h},0}^3\iiiint x^2 \Theta\left(\frac{\rho_\mathrm{net}}{\rho_\mathrm{lim}}-1\right)\mathrm{d}x\frac{\sin\theta\,\mathrm{d}\theta}{2}\frac{\mathrm{d}\phi}{2\pi}\frac{\sin\iota\,\mathrm{d}\iota}{2}\frac{\mathrm{d}\psi}{2\pi}=\\
 & \frac{4\pi}{3} d_{\mathrm{h},0}^3 \left\langle x_\mathrm{lim}^3 \right\rangle,
\end{split}
\label{eq:Veff}
\end{equation}
where I defined the dimensionless distance $x=r/d_{\mathrm{h},0}$ and its limiting value for detection at fixed sky position and inclination (which follows from Eq.\ \ref{eq:snrnet})
\begin{equation}
 x_\mathrm{lim}(\theta,\phi,\iota,\psi)=w_0(\theta,\phi,\psi)\left[1+\sum_{i=1}^{n-1}\left(\frac{w_i(\tilde\theta_i,\tilde\phi_i,\iota,\tilde\psi_i)}{w_0(\theta,\phi,\iota,\psi)}\right)^2\left(\frac{d_{\mathrm{h},i}}{d_{\mathrm{h},0}}\right)^2\right]^{1/2}.
\end{equation}
In the above expression, $(\tilde\theta_i,\tilde\phi_i)$ and $\tilde\psi_i$ represent the sky position and the polarization angle as seen by detector $i$, as opposed to $(\theta,\phi)$ and $\psi$ that pertain to the reference detector $0$. The average $\langle x_\mathrm{lim}^3 \rangle$ is over isotropic sky positions and orientations. We call such average the `geometrical factor' of the network. This is related to the `peanut factor' $f_\mathrm{p}$ defined in \citet{Chen2021} through $f_\mathrm{p}=\langle x_\mathrm{lim}^3 \rangle^{-1/3}$. For $n=1$, $\langle x_\mathrm{lim}^3 \rangle^{-1/3}=f_\mathrm{p}=2.264$ is the usual horizon-to-range ratio \citep{Finn1993,Chen2021}.

For each detector, the BNS range $\mathcal{R}$ can be defined as the radius of an Euclidean sphere whose volume is equal to the $V_\mathrm{eff}$ obtained considering only that single detector. Clearly, $\mathcal{R}_i\propto d_{\mathrm{h},i}$ \citep{Chen2021}. In particular, the relation between the horizon and the range can be expressed as $d_{\mathrm{h},0} = 2.264 \mathcal{R}_0 (m_\mathrm{c}/m_\mathrm{c,ref})^{5/6}$, where $m_\mathrm{c,ref}=1.22\,\mathrm{M_\odot}$ \citep{Chen2021}
is the reference chirp mass for which the BNS range is conventionally defined (that corresponds to an equal-mass binary with $m_1=m_2=1.4\,\mathrm{M_\odot}$).

For each pair of detectors $i$ and $j$, the probability distribution of the ratio
\begin{equation}
 \frac{w_i}{w_j} =  \frac{F_{+,i}^2(1+\cos^2\iota)^2+4 F_{\times,i}^2\cos^2\iota}{F_{+,j}^2(1+\cos^2\iota)^2+4 F_{\times,j}^2\cos^2\iota}
\end{equation}
depends only on the relative orientations of the two detectors. Samples of such distribution can be obtained in a simple way by sampling isotropic sky positions and binary orientations, computing the antenna pattern functions of the two detectors for each sampled configuration, and constructing the ratio as expressed in the above equation. The resulting samples of the ratio can then be used to compute the geometrical factor $\langle x_\mathrm{lim}^3 \rangle$ with a Monte-Carlo sum.
These facts allow us to compute the effective sensitive volume of a network to a binary of chirp mass $m_\mathrm{c}$ by knowing only the detector orientations and their BNS ranges.

Since the duty cycle of the GW detectors is not 100\%, at any time the GW detector network effectively acts as one of several possible sub-networks, depending on which combination of detectors is online. The formalism outlined above can be used to compute the effective sensitive volume of each of the sub-networks, and these can then be combined based on the fraction of time, in a given observing run, over which that particular sub-network was operating. From basic combinatorics, the number of sub-networks (i.e.\ possible combinations of online detectors) is
\begin{equation}
 N_\mathrm{c}(n)=\sum_{k=1}^{n}\left(\begin{array}{c}
                                  n\\
                                  k
                                  \end{array}\right),
\end{equation}
where the sum is over $n$-choose-$k$ Binomial coefficients.
For a 3-detector network, this is $N_\mathrm{c}(3)=3+3+1=7$. For the HLV network, these seven combinations are H, L, V, HL, LV, VH, HLV.
Let us number the observing runs of the GW detector network by an index $l$, and denote by $n_l$ the number of detectors that participated in each run. If $f_{l,j}$ is the fraction of run $l$'s time during which only the combination $j$ of detectors was online (the others being offline or presenting data quality issues), then the total effective sensitive volume of the run is
\begin{equation}
V_{l}=\sum_{j=0}^{N_\mathrm{c}(n_l)-1}f_{j,l}V_{\mathrm{eff},j,l}=\frac{4\pi}{3}d_{\mathrm{h},0,0,l}^3\sum_{j=0}^{N_\mathrm{c}(n_l)-1}f_{j,l}\left(\frac{\mathcal{R}_{0,j,l}}{\mathcal{R}_{0,0,l}}\right)^3\left \langle x_\mathrm{lim}^3\right\rangle_{j,l},
\label{eq:V_eff_run}
\end{equation}
where $\mathcal{R}_{i,j,l}$ is the BNS range of $i$-th detector of combination $j$ during run $l$, and similarly $d_{\mathrm{h},i,j,l}$ is the corresponding horizon distance. In order to obtain this expression, I made use of Eq.\ \ref{eq:Veff} and of the proportionality between horizon and range.

I note that the ratio of two effective sensitive volumes is independent of chirp mass, owing to the fact that the single-detector horizons (which are the only dimensional terms in Eq.\ \ref{eq:V_eff_run}) all share the same dependency $d_{\mathrm{h},0,0,l}\propto m_\mathrm{c}^{5/6}$. This shows that the detection rate estimate based on Eq.\ \ref{eq:P(N|N')} is insensitive to the mass distribution of the binaries of interest, as long as their S/N is reasonably well approximated by that of an inspiral of two point masses.

\subsection{Application to BNS and NSBH mergers in O4}

Equation \ref{eq:V_eff_run} allows us to write the ratio of expected numbers of BNS mergers $\mathcal{C}$ (Eq.\ \ref{eq:C_simple}) as a function of the BNS ranges of the detectors in each of the run (which we assume constant across the run for simplicity), the durations of the runs, and the sub-network time fractions $f_{j,l}$. The durations of the runs and the representative BNS ranges of the detectors that I adopted for the calculations are reported in Table \ref{tab:TR}. These are based on the BNS range plots for each run and each detector as reported in the observing run summary pages of the public Gravitational Wave Open Science (GWOSC) website\footnote{\url{https://gwosc.org/detector_status/}}, and are representative values close to the peaks of the distributions of ranges reported there. Let me note here that LVK officially divides O4 into three parts, namely O4a (May 24, 2023 to January 16, 2024), O4b (April 10, 2024 to January 28, 2025) and O4c (January 28, 2025 to November 18, 2025, according to the latest plan update). After the first nine weeks (63 d) of O4c, a long hiatus has taken place between April 1 and June 11: for that reason, I found it clearer to divide O4c into two parts, which I indicate with O4c1 (63 days from January 28 to April 1, 2025) and O4c2 (160 days from June 11 to November 18, 2025), thus removing the hiatus. In what follows, I present the results assuming this division.

For O4c, I assumed the same ranges as in O4b. In order to compute the sub-network time fractions for O1, O2 and O3, I retrieved the list of time segments that pass quality checks for the search of compact binary coalescences for each detector and each run from the GWOSC website\footnote{\url{https://gwosc.org/timeline/}}. This allowed me to extract the fraction of each run's time over which each sub-network was available. For O4, these time segment l
ists are not yet available: therefore, I estimated the sub-network time fractions using the limited information available in the GWOSC, following the method descibed in Appendix \ref{sec:fjl_model}. The result is reported in Table \ref{tab:fjl}, along with the geometrical factors computed using the ranges from Tab.\ \ref{tab:TR}.

\begin{table}
\caption{Run duration, representative BNS ranges of the detectors, and total effective sensitive time-volume to a BNS with $m_\mathrm{c}=1.22\,\mathrm{M_\odot}$ of the past GW observing runs, and projections for O5. The index $l$ is included to ease the comparison with Eq.\ \ref{eq:V_eff_run}.}
\label{tab:TR}
\centering
\begin{tabular}{ccccccc}
Index & Run & Duration  & \multicolumn{3}{c}{BNS range} & $V_{l}T_l$ \\
$l$ & ~ & (days) & \multicolumn{3}{c}{(Mpc)} & ($10^{-3}\,\mathrm{Gpc^3\,yr}$) \\
 ~ & ~ & ~ & H & L & V & ~ \\
 \hline
0 & O1  & 130  & 70 & 60 & -- & 0.43\\
1 & O2  & 268  & 60 & 85 & 25 & 1.5\\
2 & O3a & 183  & 105 & 135 & 45 & 7.7\\
3 & O3b & 147  & 115 & 135 & 50 & 7.4\\
4 & O4a & 235  & 140 & 150 & -- & 15\\
5 & O4b & 294  & 155 & 170 & 50 & 23\\
6 & O4c1 & 63  & 155 & 170 & 50 & 5.4\\
7 & O4c2 & 160  & 155 & 170 & 50 & 14\\
8 & O5  & 1000 & 330 & 330 & 150 & 940\\

\hline
 \end{tabular}

\end{table}

\begin{table*}
\caption{Fraction of past GW observing run time during which each sub-network was operational (i.e.\ was taking data that passes quality checks for the search of compact binary coalescences) and corresponding geometrical factor $\langle x_\mathrm{lim}^3\rangle$ computed using the ranges from Tab.\ \ref{tab:TR}. The indices $(j,l)$ are included to ease the comparison with Eq.\ \ref{eq:V_eff_run}.}
\label{tab:fjl}
\centering
 \begin{tabular}{cccc}

Index & Sub-network & Time fraction & $\left\langle x_\mathrm{lim}^3\right\rangle_{j,l}$ \\
$(j,l)$ & & $f_{j,l}$ & \\
\hline
\multicolumn{4}{c}{O1}\\
\hline
 (0,0) & H & 0.21$^\star$ & 0.086 \\
 (1,0) & L & 0.13$^\star$ & 0.086\\
 (2,0) & HL & 0.38 & 0.19 \\
\hline
 \multicolumn{4}{c}{O2}\\
\hline
(0,1) & H & 0.14$^\star$ & 0.086 \\
(1,1) & L & 0.125$^\star$ & 0.086 \\
(2,1) & V & 0.0062$^\star$ & 0.086 \\
(3,1) & HL & 0.38 & 0.44 \\
(4,1) & VH & 0.0064 & 0.10 \\
(5,1) & LV & 0.0083 & 0.095\\
(6,1) & HLV & 0.057 & 0.47\\
\hline
 \multicolumn{4}{c}{O3a}\\
\hline
(0,2) & H & 0.030 & 0.086 \\
(1,2) & L & 0.035 & 0.086\\
(2,2) & V & 0.086 & 0.086\\
(3,2) & HL & 0.14 & 0.36\\
(4,2) & VH & 0.096 & 0.10\\
(5,2) & LV & 0.14 & 0.097\\
(6,2) & HLV & 0.44 & 0.39\\
\hline
 \multicolumn{4}{c}{O3b (and O5)}\\
\hline
(0,3) & H & 0.031 & 0.086\\
(1,3) & L & 0.023 & 0.086\\
(2,3) & V & 0.064 & 0.086\\
(3,3) & HL & 0.16 & 0.31\\
(4,3) & VH & 0.10 & 0.11\\
(5,3) & LV & 0.093 & 0.10\\
(6,3) & HLV & 0.50 & 0.34\\
\hline
 \multicolumn{4}{c}{O4a$^{\dagger}$}\\
 \hline
(0,4) & H & 0.14 & 0.086\\
(1,4) & L & 0.16 & 0.086 \\
(2,4) & HL & 0.53 & 0.27\\
\hline
 \multicolumn{4}{c}{O4b$^{\dagger}$}\\
\hline
(0,5) & H & 0.025  & 0.086\\
(1,5) & L & 0.067  & 0.086\\
(2,5) & V & 0.080  & 0.086\\
(3,5) & HL & 0.079 & 0.28\\
(4,5) & VH & 0.093 & 0.097\\
(5,5) & LV & 0.25 & 0.095\\
(6,5) & HLV & 0.29 & 0.29\\
\hline
 \multicolumn{4}{c}{O4c$^{\dagger}$}\\
\hline
(0,6/7) & H & 0.034  & 0.086\\
(1,6/7) & L & 0.061  & 0.086\\
(2,6/7) & V & 0.080  & 0.086\\
(3,6/7) & HL & 0.093 & 0.28\\
(4,6/7) & VH & 0.12 & 0.097\\
(5,6/7) & LV & 0.22 & 0.095\\
(6,6/7) & HLV & 0.34 & 0.29\\
\hline\end{tabular}
\flushleft
\footnotesize
The KAGRA detector is not included for simplicity.\\
$^{\dagger}$Based on GWOSC detector status summary and the method described in Appendix \ref{sec:fjl_model}, because the data segments were not yet available at the time of writing.\\
$^\star$These fractions are set to zero in the computation of the effective sensitive volume of the run, because single-detector triggers were not considered valid during these runs.
\end{table*}

\begin{figure*}
 \centering
 \includegraphics[width=\textwidth]{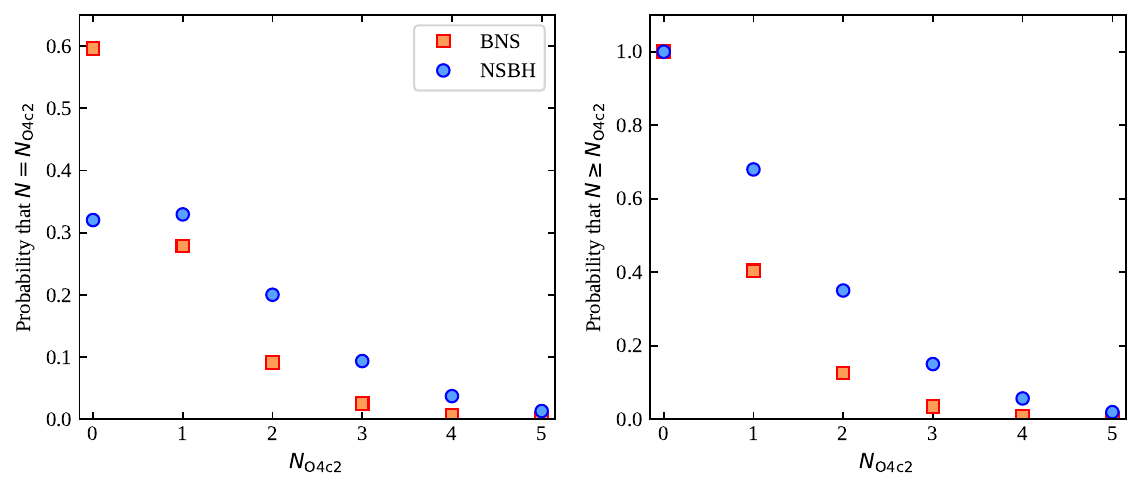}
 \caption{BNS and NSBH merger detection probability in O4c. The red squares in the left-hand panel show the probability that exactly $N_\mathrm{O4c2}$ BNS mergers are detected by the LVK network during O4c2, based on the number $N'=2$ detected in previous runs, according to Eq.\ \ref{eq:P(N|N')} and adopting the Jeffreys prior ($\alpha=1/2$). The blue circles refer to NSBH instead, assuming $N'=5$. In the right-hand panel, the probability of a number of detections $N\geq N_\mathrm{O4c2}$ in O4c2 is shown for the same two classes of sources.}
 \label{fig:PNO4c}
\end{figure*}

Using the time-volumes in Table \ref{tab:TR}, we can finally compute the  $\mathcal{C}$ ratio for some cases of interest. First of all, the ratio of the time-volume expected to be span in O4c2 with respect to that surveyed up to O4c1 is
\begin{equation}
 \mathcal{C}=\frac{VT_\mathrm{O4c2}}{VT_\mathrm{O1\to O4c1}}\approx 0.23.
\end{equation}
Assuming the number of previous detection to be $N^\prime=2$, and using Eq.\ \ref{eq:P(N|N')}, this leads to the O4c2 BNS detection probabilities shown with red squares in Figure \ref{fig:PNO4c}, adopting the Jeffreys prior (i.e.\ setting $\alpha=1/2$). While the most likely outcome is, not surprisingly, the absence of new BNS detections, the computation shows that there is still a $40\%$ probability of at least one BNS detection in O4c. For $0\leq \alpha\leq 1$, this probability varies in the range $34\%-46\%$.

The same approach can be applied to NSBH mergers, with the caveat that the inspiral-dominated signal approximation, and the fact that  cosmological effects are neglected in the calculation, can introduce some (small) systematic bias. For these sources, I assumed $N'=5$, which corresponds to the two NSBH candidates with a false alarm rate (FAR)) lower than $1/4\mathrm{yr}$ in the GWTC-3 catalog \citep{Abbott2023_GWTC3} plus the low-mass NSBH candidate GW230529\_181500 \citep{Abac2024} and the two candidates S250206dm and S241109bn released during O4 as `significant' public alerts\footnote{\url{https://gracedb.ligo.org/superevents/public/O4}} with an associated NSBH classification probability  larger than 50\%.  Again adopting the Jeffreys prior, I obtained the results shown by blue circles in Figure \ref{fig:PNO4c}, which show that the probability that O4c will yield at least one additional NSBH detection is around $68\%$, with $N=1$ being the most likely outcome (only slightly more likely than $N=0$). Adopting different priors affects the resulting probabilities by a few percent, spanning the range $64\%-71\%$ for $0\leq \alpha \leq 1$.

At any time $t$ after the start of O4c2, we can also compute the probability of at least one detection in the remainder of the run (of duration $T-t$). This is obtained from
\begin{equation}
\begin{split}
 & p\left(N>0\,|\,N',\alpha,\mathcal{C}(t)\right)=\\
 & = 1-p\left(N=0\,|\,N',\alpha,\mathcal{C}=\left[\frac{1-\frac{t}{T}}{\mathcal{C}(0)\frac{t}{T} + 1}\right]\mathcal{C}(0)\right) = \\
 & = 1-\left(\frac{\mathcal{C}(0)+1}{\mathcal{C}(0)\frac{t}{T}+1}\right)^{\alpha-1-N'}.
\end{split}
\end{equation}

\begin{figure}
 \centering
 \includegraphics[width=\columnwidth]{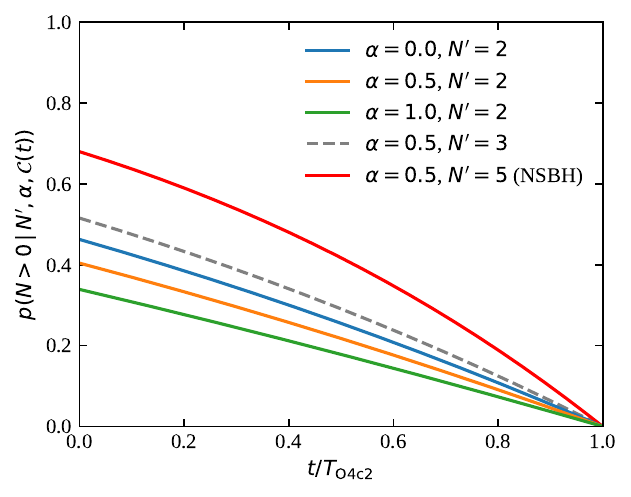}
 \caption{Probability of at least one detection in the remainder of O4, as a function of time $t$ from the start of O4c2, for three different prior choices (different colours), keeping $N'=2$ fixed (i.e.\ assuming no detection, solid lines). The dashed line represents the probability of at least one hypothetical further detection after a third detection has been made during O4c2.}
 \label{fig:PT}
\end{figure}

\noindent Solid lines in Fig.\ \ref{fig:PT} show the resulting probability for three different prior choices, $\alpha=0$, $1/2$ and $1$, keeping $N'=2$. The grey dashed line shows the result for $\alpha=1/2$ and $N'=3$, which represents the updated probability estimate of at least an additional fourth detection in the remainder of O4 after a hypothetical third detection has been made. The red solid line shows the result for NSBH mergers.

\subsection{Updated local BNS merger rate density estimate}

The time-volumes calculated with the method described in this work can be used to provide an updated estimate of the local BNS merger rate density $R_0$ based on GW observations. Using GW observational data up to the end of O3, based on the two BNS detections already discussed, and accounting for the uncertainty in the mass distribution of the merging component NSs, the LVK Collaboration estimated the true value of $R_0$ to lie in the range $10$ - $1700$ Gpc$^{-3}$ yr$^{-1}$, based on the union of the 90\% credible ranges obtained from three different methods \citep{Abbott2023_pop}. Since the merger rate density estimate scales as $R_0=N^\prime/VT$, but $N^\prime$ remained unchanged, this estimate can be updated simply by multiplying it by $VT_\mathrm{new}/VT_\mathrm{old}$. Using the time-volumes in Table \ref{tab:TR}, I concluded that the absence of BNS merger detections in O4a and O4b reduces the estimate by a factor 3.55, leading to $2.8\, \mathrm{Gpc^{-3}}\,\mathrm{yr^{-1}} \leq R_0 \leq 480\, \mathrm{Gpc^{-3}}\,\mathrm{yr^{-1}}$.
Due to the large uncertainty in models, this updated estimate remains in agreement with most predictions in the literature \citep{Mandel2022}.

\section{Predictions for O5}

\begin{figure*}
 \centering
 \includegraphics[width=\textwidth]{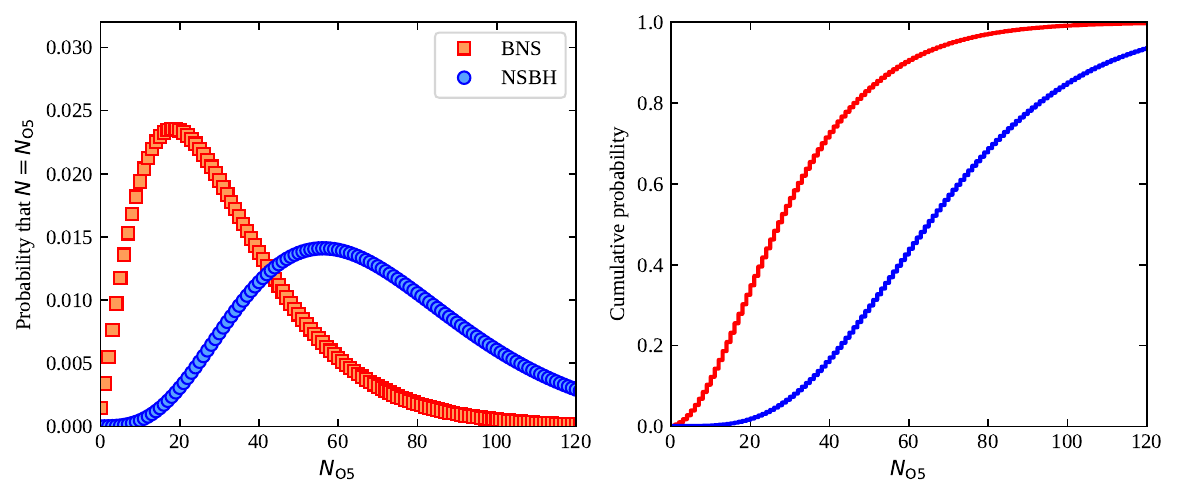}
 \caption{BNS merger detection probability in O5. The left-hand panel is the same as the corresponding panel in Fig.\ \ref{fig:PNO4c}, but for O5. The right-hand panel shows the corresponding cumulative probability.}
 \label{fig:PNO5}
\end{figure*}

During the two years that will separate the end of O4c and the next observing run O5 of the LVK network, major upgrades are anticipated to lead to a greatly improved sensitivity, with a target LIGO BNS range of 330 Mpc, and a minimum target Virgo BNS range of 150 Mpc \citep{Abbott2020LRR}. The O5 run is anticipated to last for as long as three years. Adopting the same methodology as in the previous section (still neglecting the contribution of KAGRA), with these sensitivities and run duration, and assuming the same duty cycles as for O3b, the calculation gives a time-volume to be surveyed for BNS mergers in O5 that is 12.6 times larger than the time-volume at the end of O4 (see Table . Adopting $\mathcal{C}=12.6$ and conservatively keeping $N^\prime=2$ for BNS mergers and $N^\prime=5$ for NSBH mergers, I obtained the O5 detection probabilities shown in Figure \ref{fig:PNO5}. The greatly expanded range leads to much better detection prospects than O4. Using the median and the interval comprised between the 5$^\mathrm{th}$ and 95$^\mathrm{th}$ percentiles of the cumulative probability, we can predict the number of BNS detections in O5 to be $N_\mathrm{BNS,O5}=28_{-21}^{+44}$, and the NSBH detections to be $N_\mathrm{NSBH,O5}=65_{-38}^{+61}$. These estimates are lower by more than one order of magnitude with respect to those presented in \citet{Petrov2022}, but they still demonstrate very promising prospects for multi-messenger astronomy in the next future. Clearly, they are based on a provisional estimate of the network sensitivity in O5, and hence will need to be updated once more accurate information will be available.

\section{Discussion and conclusions}

In this work I presented a relatively simple method that aims to give reliable detection rate predictions to guide the astronomical community interested in the electromagnetic follow up of BNS and NSBH mergers detected through their GWs, using publicly available information such as the BNS ranges of the detectors. The method clearly has some limitations. One of them stems from the fact that the sensitive volumes are estimated based on a simple representation of the GW detection condition as a threshold S/N. Actual search algorithms are more complex than this. In addition, the detector sensitivities typically vary during the runs. A much more accurate estimate of the surveyed time-volume can only be obtained through injection and recovery of simulated signals into the actual noise \citep[e.g.][]{Abbott2023_GWTC3}. The results of such injections in the future will allow for validating (or putting into question) the results presented here.

The results presented here support the idea that GW170817 has been a particularly lucky statistical fluctuation. With more data, we see now that the average detection rate (and consequently the rate density) of BNS mergers is not as high as we could estimate eight years ago, but this is inherent to low-number statistics. Still, the probabilities derived in this work show that the next detection is just around the corner.

\begin{acknowledgements}

Some of the content of this work appeared earlier in an LVK technical note (\url{https://dcc.ligo.org/P2400022/public}) that was made public along with the April 18, 2024 update of the LVK Public Alerts User Guide website. I acknowledge the help I received at that time from colleagues in the LVK Collaboration who participated in the internal review of that document. I also thank Francesco Pannarale for pointing out some notation errors in that document, which I corrected in this work. I thank Andrew J.\ Levan for persuading me to seek publication of this work as a scientific article. I acknowledge funding by the Italian National Institute for Astrophysics (INAF) `Finanziamento per la Ricerca' Fondamentale grant no. 1.05.23.04.04. This work has been funded by the European Union-Next Generation EU, PRIN 2022 RFF M4C21.1 (202298J7KT - PEACE).

\end{acknowledgements}

\bibliographystyle{aa}
\bibliography{references}

\begin{thebibliography}{41}
\expandafter\ifx\csname natexlab\endcsname\relax\def\natexlab#1{#1}\fi

\bibitem[{{Abac} {et~al.}(2024){Abac}, {Abbott}, {Abouelfettouh}, {Acernese},
  {Ackley}, {et~al.}}]{Abac2024}
{Abac}, A.~G., {Abbott}, R., {Abouelfettouh}, I., {et~al.} 2024, \apjl, 970,
  L34

\bibitem[{{Abbott} {et~al.}(2017{\natexlab{a}}){Abbott}, {Abbott}, {Abbott},
  {Acernese}, {Ackley}, {Adams}, {Adams}, {Addesso}, {Adhikari}, {Adya},
  {Affeldt}, {Afrough}, {Agarwal}, {Agathos}, {Agatsuma}, {Aggarwal}, {Aguiar},
  {Aiello}, {Ain}, {Ajith}, {Allen}, {Allen}, {Allocca}, {Altin}, {Amato},
  {Ananyeva}, {Anderson}, {Anderson}, {Angelova}, {Antier}, {Appert}, {Arai},
  {Araya}, {Areeda}, {Arnaud}, {Arun}, {Ascenzi}, {Ashton}, {Ast}, {Aston},
  {Astone}, {Atallah}, {Aufmuth}, {Aulbert}, {Aultoneal}, {Austin},
  {Avila-Alvarez}, {Babak}, {Bacon}, {Bader}, {Bae}, {Baker}, {Baldaccini},
  {Ballardin}, {Ballmer}, {Banagiri}, {Barayoga}, {Barclay}, {Barish},
  {Barker}, {Barkett}, {Barone}, {Barr}, {Barsotti}, {Barsuglia}, {Barta},
  {Bartlett}, {Bartos}, {Bassiri}, {Basti}, {Batch}, {Bawaj}, {Bayley},
  {Bazzan}, {B{\'e}csy}, {Beer}, {Bejger}, {Belahcene}, {Bell}, {Berger},
  {Bergmann}, {Bero}, {Berry}, {Bersanetti}, {Bertolini}, {Betzwieser},
  {Bhagwat}, {Bhandare}, {Bilenko}, {Billingsley}, {Billman}, {Birch},
  {Birney}, {Birnholtz}, {Biscans}, {Biscoveanu}, {Bisht}, {Bitossi}, {Biwer},
  {Bizouard}, {Blackburn}, {Blackman}, {Blair}, {Blair}, {Blair}, {Bloemen},
  {Bock}, {Bode}, {Boer}, {Bogaert}, {Bohe}, {Bondu}, {Bonilla}, {Bonnand},
  {Boom}, {Bork}, {Boschi}, {Bose}, {Bossie}, {Bouffanais}, {Bozzi},
  {Bradaschia}, {Brady}, {Branchesi}, {Brau}, {Briant}, {Brillet}, {Brinkmann},
  {Brisson}, {Brockill}, {Broida}, {Brooks}, {Brown}, {Brown}, {Brunett},
  {Buchanan}, {Buikema}, {Bulik}, {Bulten}, {Buonanno}, {Buskulic}, {Buy},
  {Byer}, {Cabero}, {Cadonati}, {Cagnoli}, {Cahillane}, {Bustillo},
  {Callister}, {Calloni}, {Camp}, {Canepa}, {Canizares}, {Cannon}, {Cao},
  {Cao}, {Capano}, {Capocasa}, {Carbognani}, {Caride}, {Carney}, {Diaz},
  {Casentini}, {Caudill}, {Cavagli{\`a}}, {Cavalier}, {Cavalieri}, {Cella},
  {Cepeda}, {Cerd{\'a}-Dur{\'a}n}, {Cerretani}, {Cesarini}, {Chamberlin},
  {Chan}, {Chao}, {Charlton}, {Chase}, {Chassande-Mottin}, {Chatterjee},
  {Chatziioannou}, {Cheeseboro}, {Chen}, {Chen}, {Chen}, {Cheng}, {Chia},
  {Chincarini}, {Chiummo}, {Chmiel}, {Cho}, {Cho}, {Chow}, {Christensen},
  {Chu}, {Chua}, {Chua}, {Chung}, {Chung}, {Ciani}, \&
  {Ciolfi}}]{Abbott2017_H0}
{Abbott}, B.~P., {Abbott}, R., {Abbott}, T.~D., {et~al.} 2017{\natexlab{a}},
  \nat, 551, 85

\bibitem[{{Abbott} {et~al.}(2017{\natexlab{b}}){Abbott}, {Abbott}, {Abbott},
  {Acernese}, {Ackley}, {Adams}, {Adams}, {Addesso}, {Adhikari}, {Adya}, \&
  et~al.}]{Abbott2017_GW170817}
{Abbott}, B.~P., {Abbott}, R., {Abbott}, T.~D., {et~al.} 2017{\natexlab{b}},
  \prl, 119, 161101

\bibitem[{{Abbott} {et~al.}(2017{\natexlab{c}}){Abbott}, {Abbott}, {Abbott},
  {Acernese}, {Ackley}, {et~al.}}]{Abbott2017_GW_and_GRB}
{Abbott}, B.~P., {Abbott}, R., {Abbott}, T.~D., {et~al.} 2017{\natexlab{c}},
  \apjl, 848, L13

\bibitem[{{Abbott} {et~al.}(2017{\natexlab{d}}){Abbott}, {Abbott}, {Abbott},
  {Acernese}, {Ackley}, {et~al.}}]{Abbott2017_MM}
{Abbott}, B.~P., {Abbott}, R., {Abbott}, T.~D., {et~al.} 2017{\natexlab{d}},
  \apjl, 848, L12

\bibitem[{{Abbott} {et~al.}(2018){Abbott}, {Abbott}, {Abbott}, {Acernese},
  {Ackley}, {et~al.}}]{Abbott2018_EoS}
{Abbott}, B.~P., {Abbott}, R., {Abbott}, T.~D., {et~al.} 2018, \prl, 121,
  161101

\bibitem[{{Abbott} {et~al.}(2019){Abbott}, {Abbott}, {Abbott}, {Acernese},
  {Ackley}, {et~al.}}]{Abbott2019_GWtest}
{Abbott}, B.~P., {Abbott}, R., {Abbott}, T.~D., {et~al.} 2019, \prl, 123,
  011102

\bibitem[{Abbott {et~al.}(2018)}]{Abbott2020LRR}
Abbott, B.~P. {et~al.} 2018, Living Rev. Rel., 21, 3

\bibitem[{{Abbott} {et~al.}(2023{\natexlab{a}}){Abbott}, {Abbott}, {Acernese},
  {Ackley}, {Adams}, {Adhikari}, {Adhikari}, {Adya}, {Affeldt}, {Agarwal},
  {Agathos}, {Agatsuma}, {Aggarwal}, {et~al.}}]{Abbott2023_pop}
{Abbott}, R., {Abbott}, T.~D., {Acernese}, F., {et~al.} 2023{\natexlab{a}},
  Physical Review X, 13, 011048

\bibitem[{{Abbott} {et~al.}(2023{\natexlab{b}}){Abbott}, {Abbott}, {Acernese},
  {Ackley}, {Adams}, {Adhikari}, {Adhikari}, {Adya}, {Affeldt}, {Agarwal}, \&
  et~al.}]{Abbott2023_GWTC3}
{Abbott}, R., {Abbott}, T.~D., {Acernese}, F., {et~al.} 2023{\natexlab{b}},
  Physical Review X, 13, 041039

\bibitem[{{Acernese} {et~al.}(2015){Acernese}, {Agathos}, {Agatsuma}, {Aisa},
  {Allemandou}, {et~al.}}]{Virgo2015}
{Acernese}, F., {Agathos}, M., {Agatsuma}, K., {et~al.} 2015, Classical and
  Quantum Gravity, 32, 024001

\bibitem[{{Baker} {et~al.}(2017){Baker}, {Bellini}, {Ferreira}, {Lagos},
  {Noller}, \& {Sawicki}}]{Baker2017}
{Baker}, T., {Bellini}, E., {Ferreira}, P.~G., {et~al.} 2017, \prl, 119, 251301

\bibitem[{{Chen} {et~al.}(2021){Chen}, {Holz}, {Miller}, {Evans}, {Vitale}, \&
  {Creighton}}]{Chen2021}
{Chen}, H.-Y., {Holz}, D.~E., {Miller}, J., {et~al.} 2021, Classical and
  Quantum Gravity, 38, 055010

\bibitem[{{Colombo} {et~al.}(2022){Colombo}, {Salafia}, {Gabrielli},
  {Ghirlanda}, {Giacomazzo}, {Perego}, \& {Colpi}}]{Colombo2022}
{Colombo}, A., {Salafia}, O.~S., {Gabrielli}, F., {et~al.} 2022, \apj, 937, 79

\bibitem[{{Coulter} {et~al.}(2017){Coulter}, {Foley}, {Kilpatrick}, {Drout},
  {Piro}, {Shappee}, {Siebert}, {Simon}, {Ulloa}, {Kasen}, {Madore},
  {Murguia-Berthier}, {Pan}, {Prochaska}, {Ramirez-Ruiz}, {Rest}, \&
  {Rojas-Bravo}}]{Coulter2017}
{Coulter}, D.~A., {Foley}, R.~J., {Kilpatrick}, C.~D., {et~al.} 2017, Science,
  358, 1556

\bibitem[{{Creminelli} \& {Vernizzi}(2017)}]{Creminelli2017}
{Creminelli}, P. \& {Vernizzi}, F. 2017, \prl, 119, 251302

\bibitem[{{Dhurandhar} \& {Tinto}(1988)}]{Dhurandhar1988}
{Dhurandhar}, S.~V. \& {Tinto}, M. 1988, \mnras, 234, 663

\bibitem[{{Dominik} {et~al.}(2015){Dominik}, {Berti}, {O'Shaughnessy},
  {Mandel}, {Belczynski}, {Fryer}, {Holz}, {Bulik}, \&
  {Pannarale}}]{Dominik2015}
{Dominik}, M., {Berti}, E., {O'Shaughnessy}, R., {et~al.} 2015, \apj, 806, 263

\bibitem[{{Eichler} {et~al.}(1989){Eichler}, {Livio}, {Piran}, \&
  {Schramm}}]{Eichler1989}
{Eichler}, D., {Livio}, M., {Piran}, T., \& {Schramm}, D.~N. 1989, \nat, 340,
  126

\bibitem[{{Essick}(2023)}]{Essick2023}
{Essick}, R. 2023, Physical Review D, 108, 043011

\bibitem[{{Finn} \& {Chernoff}(1993)}]{Finn1993}
{Finn}, L.~S. \& {Chernoff}, D.~F. 1993, \prd, 47, 2198

\bibitem[{{Ghirlanda} {et~al.}(2019){Ghirlanda}, {Salafia}, {Paragi},
  {Giroletti}, {Yang}, {Marcote}, {Blanchard}, {Agudo}, {An}, {Bernardini},
  {Beswick}, {Branchesi}, {Campana}, {Casadio}, {Chassande-Mottin}, {Colpi},
  {Covino}, {D'Avanzo}, {D'Elia}, {Frey}, {Gawronski}, {Ghisellini}, {Gurvits},
  {Jonker}, {van Langevelde}, {Melandri}, {Moldon}, {Nava}, {Perego},
  {Perez-Torres}, {Reynolds}, {Salvaterra}, {Tagliaferri}, {Venturi},
  {Vergani}, \& {Zhang}}]{Ghirlanda2019}
{Ghirlanda}, G., {Salafia}, O.~S., {Paragi}, Z., {et~al.} 2019, Science, 363,
  968

\bibitem[{{Kajino} {et~al.}(2019){Kajino}, {Aoki}, {Balantekin}, {Diehl},
  {Famiano}, \& {Mathews}}]{Kajino2019}
{Kajino}, T., {Aoki}, W., {Balantekin}, A.~B., {et~al.} 2019, Progress in
  Particle and Nuclear Physics, 107, 109

\bibitem[{{Kasen} {et~al.}(2017){Kasen}, {Metzger}, {Barnes}, {Quataert}, \&
  {Ramirez-Ruiz}}]{Kasen2017}
{Kasen}, D., {Metzger}, B., {Barnes}, J., {Quataert}, E., \& {Ramirez-Ruiz}, E.
  2017, \nat, 551, 80

\bibitem[{{Kasliwal} {et~al.}(2017){Kasliwal}, {Nakar}, {Singer}, {Kaplan},
  {Cook}, {Van Sistine}, {Lau}, {Fremling}, {Gottlieb}, {Jencson}, {Adams},
  {Feindt}, {Hotokezaka}, {Ghosh}, {Perley}, {Yu}, {Piran}, {Allison},
  {Anupama}, {Balasubramanian}, {Bannister}, {Bally}, {Barnes}, {Barway},
  {Bellm}, {Bhalerao}, {Bhattacharya}, {Blagorodnova}, {Bloom}, {Brady},
  {Cannella}, {Chatterjee}, {Cenko}, {Cobb}, {Copperwheat}, {Corsi}, {De},
  {Dobie}, {Emery}, {Evans}, {Fox}, {Frail}, {Frohmaier}, {Goobar}, {Hallinan},
  {Harrison}, {Helou}, {Hinderer}, {Ho}, {Horesh}, {Ip}, {Itoh}, {Kasen},
  {Kim}, {Kuin}, {Kupfer}, {Lynch}, {Madsen}, {Mazzali}, {Miller}, {Mooley},
  {Murphy}, {Ngeow}, {Nichols}, {Nissanke}, {Nugent}, {Ofek}, {Qi}, {Quimby},
  {Rosswog}, {Rusu}, {Sadler}, {Schmidt}, {Sollerman}, {Steele}, {Williamson},
  {Xu}, {Yan}, {Yatsu}, {Zhang}, \& {Zhao}}]{Kasliwal2017}
{Kasliwal}, M.~M., {Nakar}, E., {Singer}, L.~P., {et~al.} 2017, Science, 358,
  1559

\bibitem[{{Kruckow} {et~al.}(2018){Kruckow}, {Tauris}, {Langer}, {Kramer}, \&
  {Izzard}}]{Kruckow2018}
{Kruckow}, M.~U., {Tauris}, T.~M., {Langer}, N., {Kramer}, M., \& {Izzard},
  R.~G. 2018, \mnras, 481, 1908

\bibitem[{{Lazzati} {et~al.}(2017){Lazzati}, {Deich}, {Morsony}, \&
  {Workman}}]{Lazzati2017}
{Lazzati}, D., {Deich}, A., {Morsony}, B.~J., \& {Workman}, J.~C. 2017, \mnras,
  471, 1652

\bibitem[{{Li} \& {Paczy{\'n}ski}(1998)}]{Li1998}
{Li}, L.-X. \& {Paczy{\'n}ski}, B. 1998, \apjl, 507, L59

\bibitem[{{LIGO Scientific Collaboration} {et~al.}(2015)}]{LIGO2015}
{LIGO Scientific Collaboration} {et~al.} 2015, Classical and Quantum Gravity,
  32, 074001

\bibitem[{{Mandel} \& {Broekgaarden}(2022)}]{Mandel2022}
{Mandel}, I. \& {Broekgaarden}, F.~S. 2022, Living Reviews in Relativity, 25, 1

\bibitem[{{Mapelli} \& {Giacobbo}(2018)}]{Mapelli2018}
{Mapelli}, M. \& {Giacobbo}, N. 2018, \mnras, 479, 4391

\bibitem[{{Margutti} \& {Chornock}(2021)}]{Margutti2021}
{Margutti}, R. \& {Chornock}, R. 2021, \araa, 59, 155

\bibitem[{{Metzger}(2020)}]{Metzger2020LRR}
{Metzger}, B.~D. 2020, Living Reviews in Relativity, 23, 1

\bibitem[{{Mooley} {et~al.}(2018){Mooley}, {Deller}, {Gottlieb}, {Nakar},
  {Hallinan}, {Bourke}, {Frail}, {Horesh}, {Corsi}, \&
  {Hotokezaka}}]{Mooley2018}
{Mooley}, K.~P., {Deller}, A.~T., {Gottlieb}, O., {et~al.} 2018, \nat, 561, 355

\bibitem[{{Nakar}(2020)}]{Nakar2020}
{Nakar}, E. 2020, \physrep, 886, 1

\bibitem[{{Nicholl} \& {Andreoni}(2025)}]{Nicholl2025}
{Nicholl}, M. \& {Andreoni}, I. 2025, Philosophical Transactions of the Royal
  Society of London Series A, 383, 20240126

\bibitem[{{Petrov} {et~al.}(2022){Petrov}, {Singer}, {Coughlin}, {Kumar},
  {Almualla}, {Anand}, {Bulla}, {Dietrich}, {Foucart}, \&
  {Guessoum}}]{Petrov2022}
{Petrov}, P., {Singer}, L.~P., {Coughlin}, M.~W., {et~al.} 2022, \apj, 924, 54

\bibitem[{{Pian} {et~al.}(2017){Pian}, {D'Avanzo}, {Benetti}, {Branchesi},
  {Brocato}, {Campana}, {et~al.}}]{Pian2017}
{Pian}, E., {D'Avanzo}, P., {Benetti}, S., {et~al.} 2017, \nat, 551, 67

\bibitem[{{Ray} {et~al.}(2023){Ray}, {Camilo}, {Creighton}, {Ghosh}, \&
  {Morisaki}}]{Ray2023}
{Ray}, A., {Camilo}, M., {Creighton}, J., {Ghosh}, S., \& {Morisaki}, S. 2023,
  \prd, 107, 043035

\bibitem[{{Savchenko} {et~al.}(2017){Savchenko}, {Ferrigno}, {Kuulkers},
  {Bazzano}, {Bozzo}, {Brandt}, {Chenevez}, {Courvoisier}, {Diehl}, {Domingo},
  {Hanlon}, {Jourdain}, {von Kienlin}, {Laurent}, {Lebrun}, {Lutovinov},
  {Martin-Carrillo}, {Mereghetti}, {Natalucci}, {Rodi}, {Roques}, {Sunyaev}, \&
  {Ubertini}}]{Savchenko2017}
{Savchenko}, V., {Ferrigno}, C., {Kuulkers}, E., {et~al.} 2017, \apjl, 848, L15

\bibitem[{{Somiya}(2012)}]{KAGRA2012}
{Somiya}, K. 2012, Classical and Quantum Gravity, 29, 124007

\end{thebibliography}

\appendix

\section{Sub-network time fractions}\label{sec:fjl_model}

From the summary pages\footnote{The summary pages can be reached at the following urls: \url{https://gwosc.org/detector_status/O4a/}, \url{https://gwosc.org/detector_status/O4b/}} of the O4a and O4b runs on the public GWOSC website we collected the following pieces of information: the fraction $\eta_{i,l}$ of observing time of the $i$-th detector during the run, and the fraction of time $\xi_{k,l}^\mathrm{GWOSC}$ during which no interferometer ($k=0$), only one interferometer ($k=1$), two interferometers ($k=2$) or three interferometers ($k=3$) were observing (indicated as no-, single-, double- and triple-interferometer under the `network duty factor' section of each summary page). This information is not sufficient to estimate the $f_{j,l}$ fractions. To obviate to this, we assumed the following  simple model of the network activity: for a fraction $\eta_{\mathrm{cd},l}$ of the time, the network is in a coordinated downtime; for the remaining fraction $(1-\eta_{\mathrm{cd},l})$ of the run, at any time each detector is independently active with a probability
\begin{equation}
p_{\mathrm{act},i,l} = \frac{\eta_{i,l}}{1-\eta_{\mathrm{cd},l}},
\end{equation}
where the constant at the denominator ensures that the detector's total active time fraction is $\eta_{i,l}$ as expected. Conversely, the probability of the detector being inactive is
\begin{equation}
p_{\neg\mathrm{act},i,l}=1-p_{\mathrm{act},i,l}=1-\frac{\eta_{i,l}}{1-\eta_{\mathrm{cd},l}}=\frac{1-\eta_{i,l}-\eta_{\mathrm{cd},l}}{1-\eta_\mathrm{cd}}.
\end{equation}
Let us now construct the $n_l$-uple $(a_{0,j,l},a_{1,j,l},...,a_{n_l-1,j,l})$ such that $a_{i,j,l}=1$ if the $i$-th detector is active in configuration $j$ during run $l$, and $a_{i,j,l}=0$ otherwise. The sub-network time fraction predicted by the model is then
\begin{equation}
\begin{split}
 & f_{j,l} = (1-\eta_{\mathrm{cd},l})\prod_{i=0}^{n_l-1} a_{i,j,l}p_{\mathrm{act},i,l}+(1-a_{i,j,l})p_{\neg\mathrm{act},i,l} =\\
 & = (1-\eta_{\mathrm{cd},l})^{1-n_l}\prod_{i=0}^{n_l-1} a_{i,j,l}\eta_{i,l}+(1-a_{i,j,l})(1-\eta_{i,l}-\eta_{\mathrm{cd},l})
\end{split}
\label{eq:fjl_model}
\end{equation}
For example, in O4a (which corresponds to $l=4$) the configuration $j=0$ corresponds to H being active while L is inactive, that is, $(a_{0,0,4},a_{1,0,4})=(1,0)$. Then we have $f_{0,4}=\eta_{0,4}(1-\eta_{1,4}-\eta_{\mathrm{cd},4})$. Clearly, in each sub-network, the number of active detectors is
\begin{equation}
n_{\mathrm{act},j,l} = \sum_{i=0}^{n_l-1}a_{i,j,l}.
\end{equation}
This implies that, for $k\geq 1$, the model predicts
\begin{equation}
 \xi_{k,l} = \sum_{j=0}^{N_\mathrm{c}(n_l)-1}\delta_{k,n_{\mathrm{act},j,l}} f_{j,l},
\end{equation}
where $\delta_{k,n}$ is Kronecker's delta. The remaining $\xi_{0,l}$ can be obtained from the fact that $\sum_{k=0}^{n_l}\xi_{k,l}=1$. This shows that this simple model allows us to predict the fractions $\xi_{k,l}$ by specifying the single parameter $\eta_{\mathrm{cd},l}$, once the individual detector duty cycles $\eta_{j,l}$ are known. In order to choose the value of $\eta_{\mathrm{cd},l}$ that provides the best match to the reported $\xi_{k,l}^\mathrm{GWOSC}$, we minimized the sum of the squared residuals between the actual and predicted fractions,
\begin{equation}
 \Psi(\eta_{\mathrm{cd},l})=\sum_{k=0}^{n_l}\left[\xi_{k,l}(\eta_{\mathrm{cd},l})-\xi_{k,l}^\mathrm{GWOSC})\right]^2.
\end{equation}
Thid led to the values of the coordinated downtime fractions shown in Table \ref{tab:xik}, which we used to compute the sub-network time fractions reported in Table \ref{tab:fjl} using Equation \ref{eq:fjl_model}.

\begin{table}
 \caption{Fraction $\xi_{k,l}$ of run $l$'s time during which $k$ detectors were observing together, as reported for the O4a and O4b sub-runs in the GWOSC (third column) and as predicted by our simple network duty cycle model (fourth columns), assuming the fraction $\eta_{\mathrm{cd},l}$ of coordinated downtime shown in the fifth column.}
 \centering
 \label{tab:xik}
 \begin{tabular}{ccccc}
 Run & $k$ & $\xi_{k,l}^\mathrm{GWOSC}$  & $\xi_{k,l}$ & $\eta_{\mathrm{cd},l}$\\
 \hline
 O4a  & ~ & ~ & ~ & 0.13\\
 \hline
  ~      & 0 & 0.17 & 0.17\\
  ~      & 1 & 0.30 & 0.30\\
  ~      & 2 & 0.53 & 0.53\\
\hline
 O4b  & ~ & ~ & ~ & 0.10 \\
\hline
  ~      & 0 & 0.11   & 0.12\\
  ~      & 1 & 0.21   & 0.17\\
  ~      & 2 & 0.37   & 0.42\\
  ~      & 3 & 0.31   & 0.29\\
\hline
 O4c  & ~ & ~ & ~ & 0.53+0.026$^\dagger$\\
\hline
  ~      & 0 & 0.56   & 0.56\\
  ~      & 1 & 0.098  & 0.070\\
  ~      & 2 & 0.16   & 0.20\\
  ~      & 3 & 0.18   & 0.17\\
\hline
 \end{tabular}
\flushleft\footnotesize $^\dagger$We decompose the coordinated downtime of O4c into the sum of two terms, the first corresponding to the downtime related to the hiatus between April 1 and June 11, 2025 (as of June 17), and the second representing coordinated downtime in the actual periods of observation.

\end{table}

\end{document}